# Understanding the pseudocapacitance of $RuO_2$ from joint density functional theory

Cheng Zhan and De-en Jiang*

Department of Chemistry, University of California, Riverside, 92521, CA, USA

*E-mail: de-en.jiang@ucr.edu. Tel.: +1-951-827-4430

**Abstract:** Pseudocapacitors have been experimentally studied for many years in electric energy storage. However, first principles understanding of the pseudocapacitive behavior is still not satisfactory due to the complexity involved in modeling electrochemistry. In this paper, we applied a novel simulation technique called Joint Density Functional Theory (JDFT) to simulate the pseudocapacitive behavior of $RuO_2$, a prototypical material, in a model electrolyte. We obtained from JDFT a capacitive curve which showed a redox peak position comparable to that in the experimental cyclic voltammetry (CV) curve. We found that the experimental turning point from double-layer to pseudocapacitive charge storage at low scan rates could be explained by the hydrogen adsorption at low coverage. As the electrode voltage becomes more negative, H coverage increases and causes the surface structure change, leading to bended –OH bonds at the on-top oxygen atoms and large capacitance. This H coverage-dependent capacitance can explain the high pseudocapacitance of hydrous $RuO_2$. Our work here provides a first principles understanding of the pseudocapacitance for $RuO_2$ in particular and for transition-metal oxides in general.

## 1. Introduction

Capacitors are an important electronic device that can store and release electrical energy very quickly. Traditional capacitors consist of a dielectric layer sandwiched by two electrodes. By making the dielectric layer very thin, nanocapacitors have been explored both computationally and experimentally.[1-4] Due to their much larger capacity than traditional dielectric capacitors, supercapacitors store energy electrochemically and are playing important roles in electric energy storage.[5-6] Two main types of supercapacitors are used: an electric double-layer capacitor (EDLC) stores energy in the electric double layer (EDL) and a pseudocapacitor stores energy via a surface redox process. Comparing with EDLCs, pseudocapacitor has higher energy density due to the advantage of redox reaction in charge storage.[7-9] Transition metal oxide is a typical electrode material to achieve pseudocapacitive energy storage due to its multivalence oxidation states.[10-19] In particular, ruthenium oxide has been studied for many years as a promising pseudocapacitor material.[20-26]

Trasatti proposed that the charge storage mechanism of ruthenium oxide capacitor can be interpreted by the redox reaction:[21]

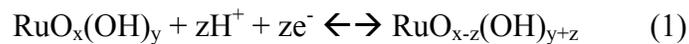

$$RuO_x(OH)_y + zH^+ + ze^- \leftrightarrow RuO_{x-z}(OH)_{y+z} \qquad (1)$$

The oxidation state of Ru can be +4, +3 and +2 during the charge storage process, which could give a maximum theoretical capacitance over 1400 F/g.[9] Experimental study on the $RuO_2$ pseudocapacitor showed that the measured capacitance from cyclic voltammetry (CV) is sensitive to the scan rate: a sharp redox peak showed up at low scan rate in the voltage of 0.1 to 0.4 V vs. reverse hydrogen electrode, while at high scan rate the CV curve is rectangle.[27-28]

Specific capacitance of ruthenium oxide can also be greatly influenced by the particle size and the amount of structure water.[28] There have been many hypotheses for explaining the capacitive behavior of ruthenium oxide,[29-30] but first principles studies have been limited. Although classical simulations and theory have been successfully applied to model EDLCs,[31-41] a first principles method such as electronic density functional theory (DFT) is needed to include the electronic structure and surface chemistry of the electrode into consideration when modeling $RuO_2$ pseudocapacitance. Although the DFT approach has been successfully applied to model the solid-state dielectric nanocapacitors,[1-4] previous DFT calculations could not quantify the pseudocapacitance of $RuO_2$ from proton adsorption and intercalation.[42-43] Directly calculating the Gibbs free energy of redox reaction could not accurately predict the electrode voltage of the reaction since it ignored the electrode-electrolyte interaction and the influence of overpotential.[44]

To be able to accurately quantify the pseudocapacitance of $RuO_2$ from first principles, one needs to take into account both the EDL and redox mechanisms at the same time. To this end, herein we employ the joint density functional theory (JDFT) that allows us to examine the electrode/electrolyte interface self-consistently by treating the electrode at the electronic-structure level and the electrolyte classically. This approach enables us to calculate the electronic chemical potential shift with the surface redox reaction and to compute the differential capacitance versus the electrode voltage, thereby providing a capacitive curve from first principles that can be directly compared with the measured pseudocapacitive behavior. In Sec. 2, we explain in detail the JDFT approach for the pseudocapacitive charge storage. Our main results are discussed in Sec. 3, and we conclude in Sec. 4.

**2. Methods**

Electronic chemical potential and total energy were calculated by Joint Density Functional Theory (JDFT) in the JDFTx package with the implementation of linear polarizable continuum model (linear PCM),[45] which has been used in several theoretical electrochemical studies before.[45-47] Periodic boundary condition (supercell) was used to describe the solid/liquid interface: here we chose the most stable and active surface of rutile $RuO_2$, the (110) surface[48-49], modeled in an orthogonal crystal with a=6.24 Å, b=12.74 Å and c=25.14 Å (2×2×1 supercell). The structural details about the $RuO_2$ crystal are provided in the supplementary data (see Figure S1 and Table S1). The space along the c direction between two periodic slabs was filled with the implicit solvation model, as shown in Figure 1a. The Generalized Gradient Approximation in the form of Perdew−Burke−Ernzerhof (GGA-PBE) functional was chosen to describe the exchange-correlation energy.[50] Ultrasoft pseudopotential was used to describe the nuclei-electron interaction.[51] The cutoff energy of plane wave basis set was 20 hartree in structure optimization and 30 hartree in electronic structure calculation. The k-points mesh for Brillouin zone sampling was 4×2×1 in structure optimization and 8×4×1 in electronic chemical potential calculation.

In our simulation, the total capacitance is defined by the total charge over the electronic chemical potential shift: $Q_{tot}/\Delta\mu$.[52-53] The total charge ($Q_{tot}$) includes both Faradaic ($Q_{ps}$) and non-Faradaic ($Q_{EDL}$) parts as shown in Figure 1b, corresponding to pseudocapacitance ($C_{ps}$) and EDL capacitance ($C_{EDL}$), respectively. Thus, it is important for us to determine how to assign $Q_{ps}$ and $Q_{EDL}$ for a given $Q_{tot}$. To solve this problem, we propose a redox-EDL competition mechanism: at very low scan rate, the electrochemical behavior is at thermodynamic equilibrium and the capacitive behavior is dominated by thermodynamic energy preference. When the electrode is charged by a certain amount such as 1 e⁻, we can compare the EDL energy ($U_{EDL}$) and hydrogen adsorption energy ($E_H$) to predict the capacitive behavior.

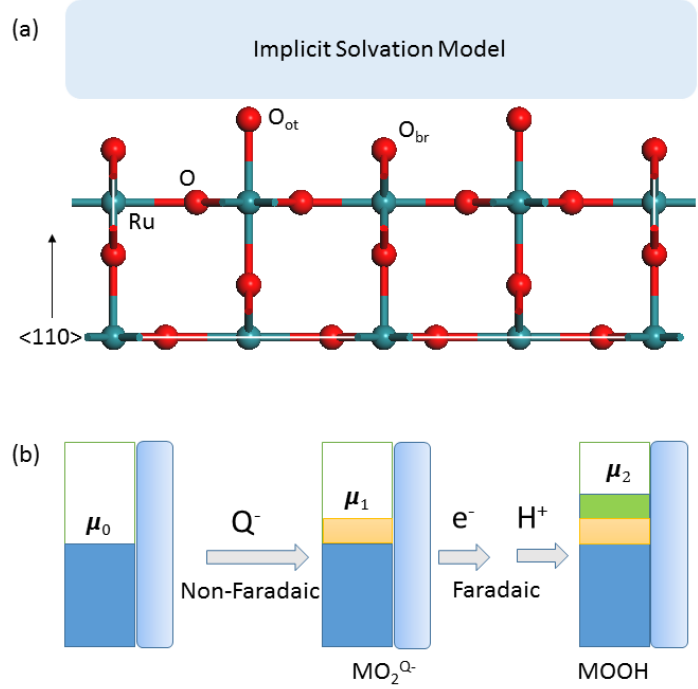

Figure 1. (a) Side view of $RuO_2$ (110). (b) Scheme of modeling the capacitance in transition metal oxide, $MO_2$.

$U_{EDL}$ is defined by the energy drop of electrode to form the EDL to neutralize the surface charge of 1 e⁻:

$$U_{EDL} = - \int_{V1}^{V2} C_{EDL} V \, dV \quad (2),$$

where $V_1$ and $V_2$ correspond to the electronic chemical potential shift with the excess charge of 1 e⁻. $E_H$, the hydrogen adsorption energy for a single H atom, is defined by

$$E_H = U_{sub+H} - U_{sub} - \frac{1}{2} U_{H2} + \Delta ZPE \quad (3),$$

where $U_{sub+H}$ and $U_{sub}$ are the total energy of H adsorbed substrate and bare substrate, respectively. ΔZPE is the difference in zero point energy of hydrogen on the substrate and that of $H_2$ molecule. We use the ΔZPE of 0.165 eV/H from previous DFT work.[42] For consecutive hydrogen adsorption, then we define $E_H(n)$, the hydrogen adsorption energy when n H atoms have been absorbed on the substrate:

$$E_H(n) = U_{sub+(n+1)H} - U_{sub+nH} - \frac{1}{2} U_{H2} + \Delta ZPE \quad (4)$$

By comparing the $U_{EDL}$ and $E_H(n)$, we can estimate which way of charge storage is more preferable to neutralize the charged surface as we progressively charge up the electrode, as indicated in Figure 2.

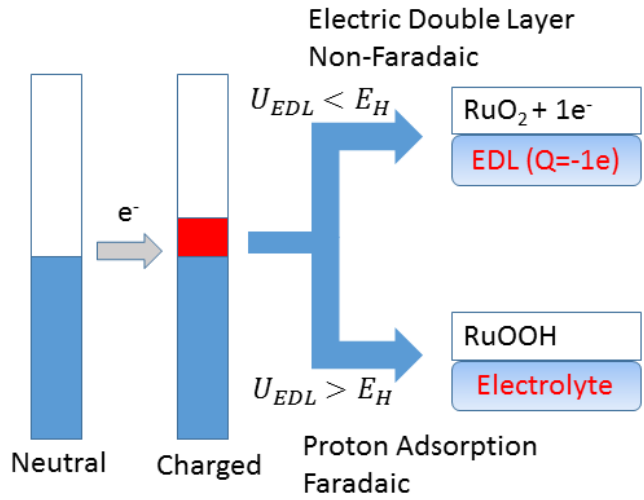

Figure 2. A brief scheme to show how to determine the capacitive behavior by comparing the $U_{EDL}$ and $E_H$.

## 3. Results and Discussion

First, we determine the potential of zero charge (PZC) and distinguish the different charge storage behavior of the anode and the cathode. Then, we focus on the capacitive behavior of the cathode. Since the EDL capacitance is always in existence as long as there is a charged surface, we will examine the intrinsic EDL capacitance of the $RuO_2(110)$ surface, and then compare the competition between EDL charge storage and redox charge storage on the cathode. Finally, based on the redox-EDL competition analysis, we will plot a capacitance-potential curve and provide a surface structure-based interpretation on the pseudocapacitive charge storage behavior of $RuO_2(110)$.

**3.1 Potential of zero charge on $RuO_2(110)$.** We need to find out the potential of zero charge on $RuO_2(110)$ in our system to assign potential region of positive and negative electrode charging behavior. For a pure $RuO_2(110)$ surface in contact with a 1M implicit electrolyte, the calculated electronic chemical potential is -4.70 eV. This calculated electronic chemical potential comes from the total consequence of work function and implicit solvation model, while the zero potential is defined far inside the electrolyte. From the JDFT benchmark calculation, the standard hydrogen electrode (SHE) is at -4.52 eV with the GGA-PBE functional.[45] Thus, our calculated potential of zero charge (PZC) is 0.18 V vs SHE. In addition, the calculated PZC is substrate thickness-independent. Experimental PZC of $RuO_2(110)$ is 0.15 V vs SHE in $10^{-3}$ M $Na_2SO_4$,[54] consistent with our DFT calculation. Here we should note that our PZC calculation is based on the electronic work function of the material with the consideration of solvation effect. The JDFTx code performs well when calculating the PZC by using linear PCM if there is no surface reaction that affects the PZC.[45] In the acidic electrolyte such as $H_2SO_4$, the PZC of transition metal oxide should be more positive so that it can resist proton adsorption to keep the surface neutral. When the electrode potential is higher than PZC, the electrode is positively charged and the capacitance is determined by the EDL because there is no available redox reaction of $SO_4^{2-}$ anion. Below the PZC, the electrode is negatively charged and the capacitance contribution is determined by the competition between the proton adsorption reaction and EDL. Thus, it is necessary to analyze the redox-EDL competition to determine the capacitive behavior of cathode.

**3.2 Influence of surface hydrogen adsorption on EDL capacitance.** Since the EDL and redox reaction could simultaneously exist when electrode is negatively charged, it is important to know how H adsorption influences the EDL capacitance. We calculated the EDL capacitance ($C_{EDL}$) of RuO$_2$(110) surface with different hydrogen coverage at various surface charge density range (Figure 3). One can see that the EDL capacitance of a pure RuO$_2$(110) is about 19 $\mu$F/cm$^2$, comparable to experimental measurement for aqueous electrolyte.[55] With the surface H adsorption, $C_{EDL}$ changed about ±2 $\mu$F/cm$^2$ within the first three H adsorption. The calculated electronic density of states (DOS) in Figure 4 also showed that H-adsorbed RuO$_2$(110) is metallic and its DOS is very similar to that of clean RuO$_2$(110), while the site-projected DOS on the H atom (Figure S2 in the supplementary data) shows that the H states are far deeper (more than 5 eV below the Fermi level). Thus, the electronic chemical potential shift is dominated by the electrolyte response (EDL capacitance). When the forth H is adsorbed on the surface, $C_{EDL}$ shows a large change and becomes very sensitive to charge. Since at the forth H adsorption, the electrode potential is close to that of hydrogen evolution on RuO$_2$, we mainly focus on the first three H adsorption.

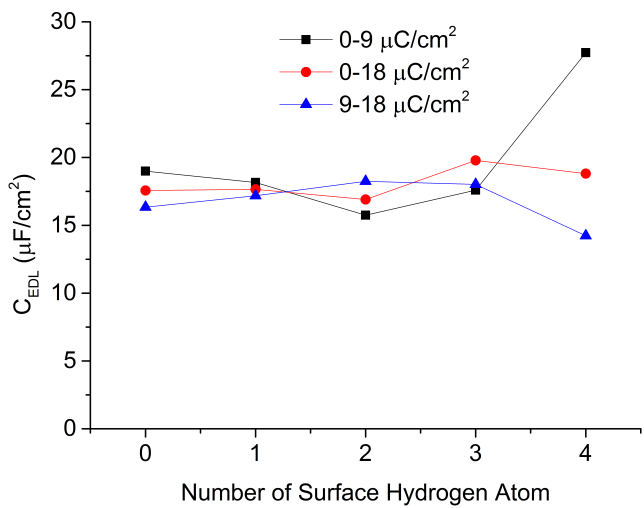

Figure 3. The electric double layer capacitance ($C_{EDL}$) at different surface H adsorption and different surface charge density range on RuO$_2$(110), based on a lateral unit cell of 79.5 Å$^2$ (6.24 Å x 12.74 Å); see Figure 7 for the lateral cell.

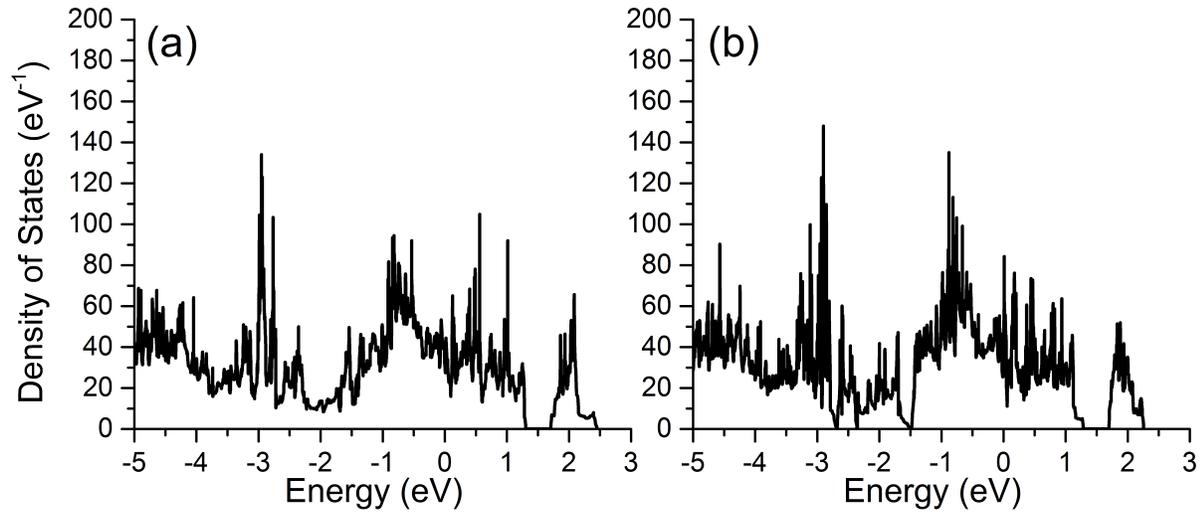

Figure 4. Total electronic density of states (DOS): (a) clean RuO$_2$(110); (b) H-adsorbed RuO$_2$(110) with 3 H atoms. See Figure 7 for the lateral cell. Fermi level is at energy zero.

**3.3 Comparison of EDL energy and H adsorption energy**. Based on the relatively constant $C_{EDL}$ of 80 F/g for the first three H adsorption, we can calculate the EDL formation energy ($U_{EDL}$) by equation (2) and compare $U_{EDL}$ with the H adsorption energy ($E_H$) to predict the capacitive behavior. With the surface charge of 1e$^-$, the charged electrode will be neutralized either by electric double layer (EDL) or H adsorption, so we plot the H adsorption energy with the EDL energy in Figure 5. One can see that H adsorption should be more preferable than the EDL formation since the H adsorption energy is more negative than $U_{EDL}$. Consequently, we conclude that the pseudocapacitive behavior of RuO$_2$(110) is dominated by the surface redox reaction for the first few proton adsorption steps when the electrode is negatively charged:

$$RuO_2 + xH^+ + xe^- \rightarrow RuO_{2-x}(OH)_x \quad (5).$$

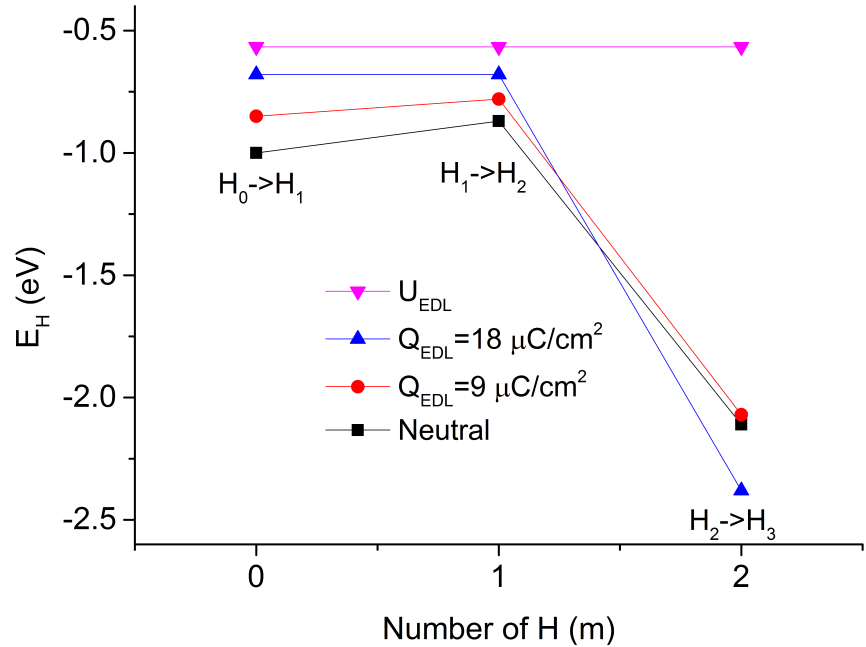

Figure 5. Comparison of the EDL formation energy ($U_{EDL}$; top line) and H adsorption energies at three different EDL charges (bottom three lines) for the three H-adsorption steps on $RuO_2(110)$. See Figure 7 for the lateral cell.

**3.4 Capacitive behavior on $RuO_2(110)$.** Now that we have found out the PZC and the preferred capacitive mechanism below PZC, the overall capacitive behavior should be: below PZC, redox reaction based on H adsorption; above PZC, EDL charge storage. The calculated charge-voltage (Q-V) and capacitance-voltage (C-V) curves are plotted in Figure 6. The calculated PZC is 0.18 V vs SHE in the Figure 6a. In the redox reaction, each proton could provide the charge density of 10.07 $\mu$C/cm$^2$.

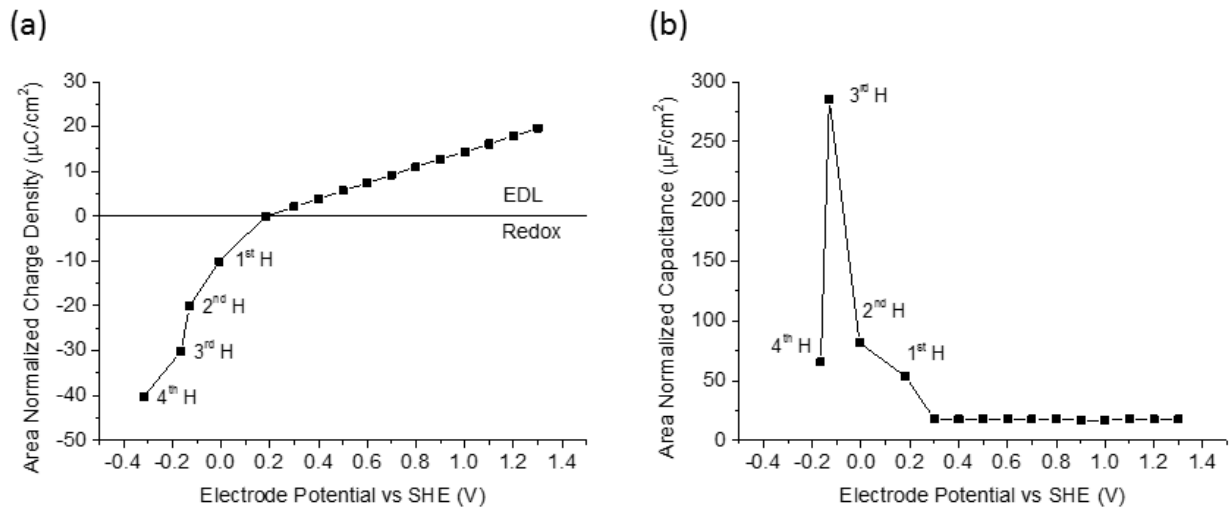

Figure 6. (a) Charge vs. voltage curve of $RuO_2$(110) in 1M LPCM electrolyte. (b) Capacitance vs. voltage curve corresponding to (a), obtained from numerical differentiation at each voltage point based on the charges at this point and the point on the right. Hydrogen adsorption sequence is labeled in both curves. Specific capacitance in F/g is provided in Figure S3 based on $RuO_2$ nanoparticles of 10 nm in size and bulk density of 6.97 g/cm$^3$.[56] The uncertainty of the differential capacitance is about ±50$\mu$F/cm$^2$ below 0.2 V and ±2$\mu$F/cm$^2$ above 0.2 V.

We calculated the electronic chemical potential shift after the redox reaction and the capacitance by C=Q/Δ$\mu$. The pseudocapacitive region in the experimental CV curve is 0.1 – 0.4 V vs SHE, corresponding to the -0.1 – 0.2 V in our simulated CV curve in Figure 6b, since we cannot take into account the influence of pH on the PZC calculation due to the limitation of implicit solvation model. The first two H adsorption could give the potential drop about 0.3 V, which exactly matched the pseudocapacitive region in the experimental CV.[28] Consequently, we propose that the experimentally measured pseudocapacitance of $RuO_2$ at 0.1 – 0.4 V vs SHE is contributed by the first two proton adsorption reaction in our simulation and the turning point in CV is due to the charge storage mechanism change from electric double layer to redox reaction at the PZC. The calculated areal capacitance is about 50-80 $\mu$F/cm$^2$. It is expected that the capacitance will have a large increase when the electrode potential is more negative (still higher than the hydrogen evolution reaction potential), corresponding to the third H adsorption on the surface. The calculated capacitance for the third H adsorption (the slope of 2H and 3H in Figure 6a) is over 260 $\mu$F/cm$^2$. This very high capacitance is due to the surface phase transition and could be a reason to explain the extremely high capacitance in hydrous RuO$_2$ nanoparticle, which will be discussed below. For the forth H adsorption on the surface, the calculated capacitance (the slope of 3H and 4H in Figure 6a) decreases to 66 $\mu$F/cm$^2$; here we note that the potential is negative enough (-0.4 vs. SHE) that hydrogen evolution reaction may occur.[57]

**3.5 Adsorption structure of hydrogen on RuO$_2$(110) and its influence on pseudocapacitance**. Now we analyze the adsorption structures of H on RuO$_2$ (110) for the redox mechanism below PZC. For the first H adsorption step (Figure 7a), the adsorption energy is about -1.0 eV both on O$_{ot}$ (on top) and O$_{br}$ (bridge) indicating that these two types of surface oxygen atoms show similar reactivity. The structure optimization after the second H adsorption also gave a perpendicular OH bond to surface (Figure 7b) and H adsorption energy of -0.87 eV. The perpendicular OH bond on RuO$_2$(110) has been experimentally observed in basic condition.[58] The first and second hydrogen adsorption yielded a capacitance of 53.3 $\mu$F/cm$^2$ (slope of 0H and 1H in Figure 6a) and 81.9 $\mu$F/cm$^2$ (slope of 1H and 2H in Figure 6a), respectively. When the third proton is adsorbed on the surface (Figure 7c), it pushed the perpendicular OH bond on O$_{ot}$ to bended OH bond and caused the surface phase transition, but there was negligible change to the underlying RuO$_2$(110) structure. Thus, the third H adsorption energy is below -2.0 eV, which is much lower than the first two and closer to the previous DFT work.[42] The corresponding capacitance caused by the third H adsorption is 266 $\mu$F/cm$^2$ (slope of 2H and 3H), an unexpectedly high value. When we put the forth H atom on the surface, the forth OH bond is perpendicular to the surface after structure optimization (Figure 7d), and the corresponding capacitance is 66 $\mu$F/cm$^2$ (slope of 3H and 4H). Consequently, we conclude that the capacitance provided by H adsorption reaction is structure-dependent: the perpendicular OH bond formation gives the capacitance about 60 $\mu$F/cm$^2$ and the capacitance from bended OH bond formation is over 200 $\mu$F/cm$^2$. This conclusion could be a possible interpretation of the experimentally observed size-dependent pseudocapacitance of RuO$_2$ nanoparticle and high capacitance of hydrous RuO$_2$.[27-28] When breaking down the crystal to small nanoparticles, it could have more O$_{ot}$ atoms that can form the bended OH bond, which can have higher capacitance than ideal surface. Hydrous RuO$_2$ has had some H atoms and OH bond on the surface, thus it could directly form bended OH bond instead of perpendicular OH bond at low H coverage.

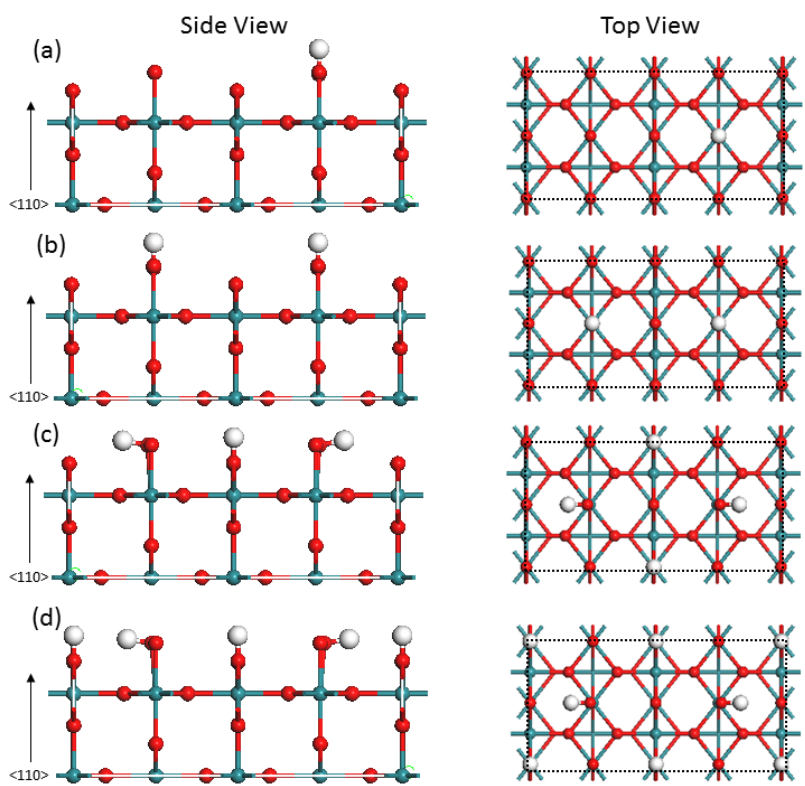

Figure 7. Side and top views of H adsorption on the RuO$_2$ (110) surface: (a) 1 H; (b) 2 H; (c) 3H; (d) 4 H atoms in a lateral unit cell (dotted lines) of 79.5 Å$^2$ (6.24 Å x 12.74 Å). Color code: Ru, cyan; O, red; H, white.

Recently, Watanabe et al. investigated the interfacial structure of $RuO_2$(110) and water in different electrochemical conditions from first principles,[59] and found that the orientation of OH bond and the surface coverage ratio of proton are strongly dependent on the pH and electrode potential. Our simulation was limited by the implicit solvation model where the ion is considered to be a point charge in the Debye screening theory and the solvent is a continuum dielectric medium with an electron density-dependent local dielectric constant. In this electrode/electrolyte system, the JDFT method considers only the electrostatic interaction between the electrode and the electrolyte, so the ion/solvent interaction and hydration are not included, but the $H^+$ ion is explicitly included via formation of the surface OH groups. In addition, the JDFT approach is intended to study the equilibrium properties, so the transport behavior such as diffusion cannot be examined. More advanced methods such as the effective-screening-medium method[60] can address the solvation and transport factors by including both the explicit solvation model and the ion dynamics. With such model, we expect that the bending angle of Ru-O-H would be influenced by the hydrogen bonding with water molecules in the electrolyte. Nevertheless, the present work provides a reasonable interpretation of the pseudocapacitive peak in the experimental CV curve of $RuO_2$:[28] the steep increase of capacitance (or current) at 0.4 V in low scan rates is caused by the charge storage mechanism changing from EDL to H adsorption reaction.

The present work focused on the (110) surface of $RuO_2$. The experimentally synthesized $RuO_2$ film has (110), (101), (100), and (1010) facets: although these surfaces have close surface energies from DFT calculation, the surface energy of (110) is the lowest.[48] That is why we chose this surface. It would be interesting to examine the pseudocapacitance on other surfaces as well.

We expect that the specific values of capacitance would change for those different surfaces, but the general idea of coverage dependence and OH formation would still apply.

**4. Summary and conclusions**

Based on the Joint Density Functional Theory with an implicit solvation model, we calculated the pseudocapacitive behavior of the $RuO_2$(110) surface. Our calculated pseudocapacitance and corresponding voltage region show a qualitative agreement with the experimental CV curve, thereby providing a reasonable interpretation on the capacitive behavior of $RuO_2$. When electrode potential is above PZC, measured capacitance is dominated by the electric double layer capacitance. Below the PZC, the capacitance is contributed by H adsorption reaction with low surface H coverage and perpendicular OH bond, which giving the capacitance of ~60 $\mu F/cm^2$. As the electrode potential goes to more negative, surface phase transition happens due to the increasing H coverage and the bended OH bond formation produces a very high capacitance over 200 $\mu F/cm^2$. These different capacitances of perpendicular and bended OH bond could explain why the small $RuO_2$ nanoparticle and hydrous $RuO_2$ have much higher capacitance than the $RuO_2$ film. Hence, our theoretical investigation provides an understanding of the pseudocapacitive charge storage behavior of the $RuO_2$ surface.

**Acknowledgement:**

This research is sponsored by the Fluid Interface Reactions, Structures, and Transport (FIRST) Center, an Energy Frontier Research Center funded by the U.S. Department of Energy (DOE), Office of Science, Office of Basic Energy Sciences. We thank Dr. R. Sundararaman for help in using the JDFTx code and Prof. Jianzhong Wu for helpful discussion. This research used resources of the National Energy Research Scientific Computing Center, a DOE Office of

Science User Facility supported by the Office of Science of the U.S. Department of Energy under Contract No. DE-AC02-05CH11231.